\documentclass[pra,showpacs,nofootinbib,longbibliography,notitlepage]{revtex4-1}
\usepackage{hyperref}
\usepackage{amsmath}
\usepackage{amssymb}
\usepackage{amsthm}
\usepackage{graphics}
\usepackage{graphicx}
\usepackage{mathtools}
\usepackage{color}
\usepackage{soul}  

\usepackage{lineno}

\newcommand*\patchAmsMathEnvironmentForLineno[1]{%
  \expandafter\let\csname old#1\expandafter\endcsname\csname #1\endcsname
  \expandafter\let\csname oldend#1\expandafter\endcsname\csname end#1\endcsname
  \renewenvironment{#1}%
     {\linenomath\csname old#1\endcsname}%
     {\csname oldend#1\endcsname\endlinenomath}}%
\newcommand*\patchBothAmsMathEnvironmentsForLineno[1]{%
  \patchAmsMathEnvironmentForLineno{#1}%
  \patchAmsMathEnvironmentForLineno{#1*}}%
\AtBeginDocument{%
\patchBothAmsMathEnvironmentsForLineno{equation}%
\patchBothAmsMathEnvironmentsForLineno{align}%
\patchBothAmsMathEnvironmentsForLineno{flalign}%
\patchBothAmsMathEnvironmentsForLineno{alignat}%
\patchBothAmsMathEnvironmentsForLineno{gather}%
\patchBothAmsMathEnvironmentsForLineno{multline}%
}

\long\def\ca#1\cb{}

\def\bra#1{\langle#1|}

\def\ket#1{|#1\rangle }

\def\Tr#1{\textrm{Tr}(#1)}
\def\pTr#1#2{\textrm{Tr}_{#1}(#2)}

\def\AC{{\cal A}}

\def\CC{{\cal C}}

\def\EC{{\cal E}}

\def\HC{{\cal H}}

\def\LC{{\cal L}}

\def\QC{{\cal Q}}
\def\RC{{\cal R}}
\def\SC{{\cal S}}

\newtheorem{thm1}{Theorem}

\newtheorem{lem1}[thm1]{Lemma}

\newtheorem{innercustomthm}{Theorem}
\newenvironment{customthm}[1]
  {\renewcommand\theinnercustomthm{#1}\innercustomthm}
  {\endinnercustomthm}

\begin{document}
\title{General approach to quantum channel impossibility by local operations and classical communication}
\author{Scott M. Cohen}
\email{cohensm52@gmail.com}
\affiliation{Department of Physics, Portland State University, Portland OR 97201}

\begin{abstract}
We describe a general approach to proving the impossibility of implementing a quantum channel by local operations and classical communication (LOCC), even with an infinite number of rounds, and find that this can often be demonstrated by solving a set of linear equations. The method also allows one to design an LOCC protocol to implement the channel whenever such a protocol exists in any finite number of rounds. Perhaps surprisingly, the computational expense for analyzing LOCC channels is not much greater than that for LOCC measurements. We apply the method to several examples, two of which provide numerical evidence that the set of quantum channels that are not LOCC is not closed and that there exist channels that can be implemented by LOCC either in one round or in three rounds that are on the boundary of the set of all LOCC channels.

Although every LOCC protocol must implement a separable quantum channel, it is a very difficult task to determine whether or not a given channel is separable. Fortunately, prior knowledge that the channel is separable is not required for application of our method. 
\end{abstract}

\date{\today}

\maketitle

Quantum information theory is concerned with the study of the transmission, storage, and manipulation of information when that information is encoded in the form of quantum systems. As it is impossible to completely isolate these systems, any description of their evolution must take into account their interaction with an environment, suggesting an analysis utilizing the powerful techniques available for the study of open quantum systems \cite{FeynmanVernon,Stinespring,Kraus,Choi}. Interaction with the environment is then considered to introduce noise into the system's evolution, and there are many important questions one may ask about the action of noisy quantum operations, or channels, not the least of which is to find the capacity of a noisy channel to transmit quantum information \cite{DevetakQCap,LloydQCap,ShorQCap}. When the input to the channel is one part of an entangled state, this capacity also measures its ability to establish entanglement between sender and receiver \cite{Schumacher,BennettQErase}.

For the input to be part of an entangled state, it must have first undergone an entangling evolution of its own. Hence, one is led to investigate the entangling capabilities of such evolutions, be they unitary or general quantum operations. The relatively recent recognition of the practical value of entangled states---examples ranging from teleportation \cite{BennettTele} and superdense coding \cite{BennettDense} to quantum cryptography \cite{Wiesner,BB84,Ekert} and quantum computation \cite{FeynmanQComp,Deutsch,NielsenChuang}---has generated considerable interest in understanding the conditions under which entanglement can be created. Thus, one may ask what characterizes a (multipartite) quantum channel's ability to create and/or increase entanglement between the subsystems upon which it acts. One significant such characterization is that the entanglement cannot increase when the channel can be simulated by local quantum operations and classical communication (LOCC), which are the only operations that can actually be implemented by spatially separated parties who lack the means to bring their subsystems together in a single laboratory. This result provides an important connection between the study of LOCC and that of entanglement. We note that it is always possible to implement any channel by LOCC when enough entanglement is pre-shared between the various parties---by using teleportation \cite{BennettTele} or by perhaps more efficient means \cite{ourNLU,myoptNLU,StahlkeU}---indicating that entanglement is an important resource \cite{HoroRMP,Aberg,PlenioResource} under the restriction to LOCC.

The importance of LOCC in quantum information processing has long been recognized, it playing a key role in teleportation \cite{BennettTele}, entanglement distillation \cite{BennettMixedQEC,BennettPurifyTele}, one-way \cite{RaussendorfBriegel} and distributed \cite{CiracDistComp} quantum computing, local cloning \cite{ourLocalCloning}, quantum secret sharing \cite{HillerySecret,VladSecret}, and beyond. While many important results have been obtained concerning LOCC \cite{IBM_CMP,NisetCerf,myLDPE,mySEPvsLOCC,myMany,KKB,ChildsLeung,ChitHsieh1,ChitDuanHsieh,FuLeungManciska,ChitHsieh2,FortescueLo,Chitambar,ChitCuiLoPRL,ChitCuiLoPRA,WinterLeung}, it has nonetheless proven difficult to characterize in simple terms. Recently \cite{myLOCCbyFirstMeas}, we presented a method of designing LOCC protocols for quantum measurements that succeeds for every measurement that can be implemented by finite-round LOCC, and for which failure has the immediate implication that the measurement cannot be implemented by LOCC no matter how many rounds of communication are allowed, including when the number of rounds is infinite (see also \cite{mySEPvsLOCC,myMany} for a different approach to designing LOCC protocols for measurements, an approach which is, in contrast, restricted to a finite number of rounds). 

There are, of course, significant differences between channels and measurements. A noteworthy example in the context of LOCC is the finding in \cite{WinterLeung} that the set of finite-round LOCC channels has a non-empty interior, whereas the interior of the set of finite-round LOCC measurements is, in fact, empty \cite{myFinInfRnds}. It is also known \cite{ChitCuiLoPRL} that the set of LOCC channels is not closed, even for the simplest possible case of two qubits \cite{WinterLeung}, but the question of whether or not LOCC measurements is a closed set appears to be as yet unanswered.

Another key difference between measurements and channels is that there are many measurements that are all associated with any given channel, see Eq.~\eqref{eqn21} below. Therefore, one can demonstrate that a channel is LOCC by showing that any one of those measurements is itself LOCC. If, on the other hand, a given channel is not LOCC, \emph{none} of the measurements associated with it can be implemented by LOCC. This latter point may lead one to expect that the characterization of LOCC channels may be a far more difficult task than that for LOCC measurements. It turns out, however that this expectation is overly pessimistic. In the present paper, we show that the method of \cite{myLOCCbyFirstMeas} for measurements is readily extended to the case of multipartite quantum channels, allowing one to design an LOCC protocol to implement a given channel whenever this is possible in a finite number of rounds, with a computational effort for the case of channels that is never more than a quadratic increase over that needed for measurements. Our method builds a protocol round by round, starting with the first one. If at any point in this process, the method finds that no measurement at the `next' round is possible, then this implies that no LOCC protocol exists for the desired channel, even with an infinite number of rounds \cite{myLOCCbyFirstMeas}. As we will see below, if no measurement is possible at the very first round, then the method proves LOCC-impossibility via the solution of a set of linear equations. We will demonstrate that solving this set of linear equations is often sufficient for proving LOCC-impossibility, illustrating this approach with various examples.

The remainder of the paper is organized as follows. We begin by reviewing the Kraus representation of quantum channels and the concomitant unitary freedom in such representations. Following this, we give a more detailed explanation of what exactly LOCC means, and how it can be represented. Then, we recall a lemma from \cite{myLOCCbyFirstMeas}, and show how the method presented there can be fully adapted to the present case of quantum channels. These ideas are then used to analyze certain example channels which are, with relative ease, found to be impossible by LOCC. Finally, we offer our conclusions.

Each quantum channel may be described by a set of Kraus operators $\{K_i\}_{i=1}^N$ \cite{Kraus}, which indicate how that channel transforms the state of the quantum system upon which it acts. If that system, described by Hilbert space $\HC$, starts out in state $\rho_0$, then its final state will be given by
\begin{align}\label{eqn19}
\rho_f = \sum_{i=1}^NK_i\rho_0K_i^\dag,
\end{align}
where with $I_\HC$ the identity operator on $\HC$,
\begin{align}\label{eqn20}
\sum_{i=1}^NK_i^\dag K_i=I_\HC
\end{align}
so that the channel is trace-preserving, $\Tr{\rho_f}=\Tr{\rho_0}$. Of course as is well-known, the set of Kraus operators describing any given channel is far from unique. For any other set of Kraus operators, $\{K_j^\prime\}_{j=1}^{N^\prime}$, that describes the same channel as the original set, let $\hat N=\max(N,N^\prime)$ and pad the smaller of the two sets with additional zero operators so that the two sets have the same number of members. Then, there exists a unitary matrix $V$ such that for each $j=1,\cdots,\hat N$,
\begin{align}\label{eqn21}
K^\prime_j=\sum_{i=1}^{\hat N}V_{ji}K_i.
\end{align}
Furthermore, any other set describing this channel may be expressed in this way \cite{NielsenChuang}.

An LOCC protocol involves one party making a measurement, informing the other parties of her outcome, after which according to a pre-approved plan, the other parties know who is to measure next and what that measurement should be. Since each local measurement involves a number of possible outcomes, the entire process is commonly represented as a tree, the children of any given node representing the set of outcomes of the measurement made at that stage in the protocol. Consider the cumulative action of all parties up to a given node $n$ in the tree, represented by Kraus operator $K$, which since the parties each make only local measurements, must be of the product form, $A\otimes B\otimes\cdots$. We will label each such node by the positive operator $K^\dag K$, commonly referred to as a \emph{POVM element}, and we will say that node $n$ is `equal' to its label $K^\dag K$. For any such tree, the root node represents the situation present before any party has yet measured and will therefore always be equal to the identity operator $I_\HC=I_A\otimes I_B\otimes\cdots$, where $I_\alpha$ is the identity operator on the local Hilbert space $\HC_\alpha$ for each party $\alpha$. At the end of any finite branch of the tree is a leaf node (leaf nodes being those that do not themselves have children), which since it is terminal, must be labeled by a POVM element associated with a Kraus representation of the desired channel. Similarly, any infinite branch has a sequence of nodes that approach such a POVM element in the limit. When we say that a finite-round LOCC protocol implements Kraus operators $\{K_j\}$ this means that each leaf is proportional to one of the $K_j^\dag K_j$, with positive constant of proportionality, and the sum of all leaves proportional to $K_j^\dag K_j$ is exactly $K_j^\dag K_j$. For infinite protocols, it means the same but in the limit as the number of rounds approaches infinity. Note that the class of infinite-round protocols can be divided into two distinct sub-classes \cite{WinterLeung}. The first subclass, which is a subset of what we will refer to as `LOCC' \cite{WinterLeung}, involves sequences of protocols more and more closely approaching a given channel simply by adding more and more rounds of communication, but without changing the local measurements implemented in earlier rounds. The second subclass, $\overline{\textrm{LOCC}}$ in \cite{WinterLeung}, includes limits of sequences in which measurements made at the earlier rounds are changed from one protocol in the sequence to the next. As indicated by the notation, $\overline{\textrm{LOCC}}$ is the topological closure of LOCC, and as mentioned above when considering quantum channels, $\overline{\textrm{LOCC}}\ne$ LOCC \cite{ChitCuiLoPRL}. We note that the results of the present paper apply to the entire class LOCC, including those that involve an infinite number of rounds, but not to $\overline{\textrm{LOCC}}$.

We will need the following lemma, proved in \cite{myLOCCbyFirstMeas} (but reworded here to conform to the present context).
\begin{lem1}\label{lem1}
Suppose the tree $\LC$ represents an LOCC protocol (finite or infinite), which implements the set of Kraus operators $\{K_j^\prime\}$. Then the POVM element $E$, representing the accumulated action of all parties up to any given node in $\LC$, is equal to a positive linear combination of the set of operators, $\{K_j^{\prime\dag}K_j^\prime\}$. That is, $E=\sum_jc_jK_j^{\prime\dag}K_j^\prime$, with $c_j\ge0$.
\end{lem1}

Without loss of generality, let us assume that Alice measures first to initiate an LOCC protocol that implements quantum channel $\EC$. More specifically, assume the protocol implements any set of Kraus operators, $\{K_j^\prime\}$, that represents $\EC$. Since Alice has measured first, the other parties have as yet done nothing, which implies that the outcome of Alice's initial measurement will correspond to a (multipartite) Kraus operator of the form $A\otimes I_{\bar A}$, where $I_{\bar A}$ is the identity operator on $\HC_{\bar A}$, the Hilbert space describing all parties other than Alice. The associated POVM element is then $E=\AC\otimes I_{\bar A}$, with $\AC=A^\dag A$, which according to Lemma~\ref{lem1} implies that with $c_j\ge0$,
\begin{align}\label{eqn22}
\AC\otimes I_{\bar A} &= \sum_{j=1}^{\hat N} c_j K_j^{\prime\dag} K_j^\prime = \sum_{i,i^\prime=1}^{\hat N}\left[\sum_{j=1}^{\hat N}V_{ji}^\ast c_jV_{ji^\prime}\right]K_i^\dag K_{i^\prime}\notag\\
				&=\sum_{i,i^\prime=1}^{\hat N}\CC_{ii^\prime}K_i^\dag K_{i^\prime},
\end{align}
and $\CC_{ii^\prime} = \sum_{j=1}^{\hat N}V_{ji}^\ast c_jV_{ji^\prime}$.

At this point it is convenient to choose a set of index pairs, $\SC\subseteq\{(i,i^\prime)\}_{i,i^\prime=1}^{\hat N}$, such that the set of operators $\{K_i^\dag K_{i^\prime}\}_{(i,i^\prime)\in\SC}$ is linearly independent. Then with $K_j^\dag K_{j^\prime}=\sum_{(i,i^\prime)\in\SC}r_{jj^\prime}^{ii^\prime}K_i^\dag K_{i^\prime}$ for $(j,j^\prime)\not\in\SC$,
\begin{align}\label{eqn29}
\AC\otimes I_{\bar A} &= \sum_{(i,i^\prime)\in\SC}\left(\CC_{ii^\prime}K_i^\dag K_{i^\prime} + \sum_{(j,j^\prime)\not\in\SC}\CC_{jj^\prime}r_{jj^\prime}^{ii^\prime}K_i^\dag K_{i^\prime}\right)\notag\\
				&=\sum_{(i,i^\prime)\in\SC}\tilde\CC_{ii^\prime}K_i^\dag K_{i^\prime},
\end{align}
where $\tilde\CC_{ii^\prime}=\CC_{ii^\prime} + \sum_{(j,j^\prime)\not\in\SC}\CC_{jj^\prime}r_{jj^\prime}^{ii^\prime}$. This condition must hold for all sets of Kraus operators $\{K_i\}$ representing $\EC$. Since Kraus representation $\{K_j^\prime\}$ for $\EC$ can also be arbitrarily chosen, we see that Eq.~\eqref{eqn29} with $0\le\AC\not\propto I_A$ (or else Alice did not actually measure) is a necessary condition for the existence of an LOCC protocol that implements $\EC$.

Let $\{\Lambda_\mu\}$ and $\{\Gamma_\nu\}$ be orthonormal bases of the space of operators acting on $\HC_A$ and $\HC_{\bar A}$, respectively, $\Tr{\Lambda_\mu^\dag\Lambda_{\mu^\prime}}=\delta_{\mu\mu^\prime}$ and $\Tr{\Gamma_\nu^\dag\Gamma_{\nu^\prime}}=\delta_{\nu\nu^\prime}$, with $\Lambda_0=I_A$ and $\Gamma_0=I_{\bar A}$. Choose any $\mu$ and any $\nu\ne0$ and multiply Eq.~\eqref{eqn29} by $\Lambda_\mu^\dag\otimes\Gamma_\nu^\dag$ to obtain
\begin{align}\label{eqn24}
0=\sum_{(i,i^\prime)\in\SC}\tilde\CC_{ii^\prime}\Tr{\left[\Lambda_\mu^\dag\otimes\Gamma_\nu^\dag\right]K_i^\dag K_{i^\prime}}=\sum_{(i,i^\prime)\in\SC}\tilde\CC_{ii^\prime}\QC_{i^\prime i}^{(\mu\nu)},~\mu=0,\cdots,d_A^2-1,~\nu=1,\cdots,d_{\bar A}^2-1
\end{align}
where $d_A$ is the dimension of $\HC_A$, $d_{\bar A}$ is that of $\HC_{\bar A}$, and
\begin{align}\label{eqn25}
\QC_{i^\prime i}^{(\mu\nu)}:=\Tr{\left[\Lambda_\mu^\dag\otimes\Gamma_\nu^\dag\right]K_i^\dag K_{i^\prime}}.
\end{align}

The next step is to form the coefficients $\tilde\CC_{ii^\prime}$ into an $\vert\SC\vert$-dimensional column vector $\vec c$ (by, say, stacking each column one below the next) and also the $\QC_{i^\prime i}^{(\mu\nu)}$ into a row vector for each $\mu$ and $\nu\ne0$, collecting all the latter row vectors into a matrix ${\bf Q}_{_A}$. Then, we have from Eq.~\eqref{eqn24},
\begin{align}\label{eqn28}
0={\bf Q}_{_A}\vec c,
\end{align}
showing that any initial local measurement by Alice in an LOCC protocol exactly implementing the desired quantum channel $\EC$ is determined by the nullspace of matrix ${\bf Q}_{_A}$.\footnote{Note that once the index set $\SC$ and the orthonormal bases $\{\Lambda_\mu\}$ and $\{\Gamma_\nu\}$ are chosen, matrix ${\bf Q}_A$ is completely defined by the channel itself through the Kraus representation $\{K_i\}$. In addition, we claim that the choice of index set and bases does not change the results of our approach. The reason why the basis choices change nothing has been discussed in footnote $4$ of \cite{myLOCCbyFirstMeas}. Choosing index set $\SC^\prime$ instead of $\SC$, corresponds to a choice of linearly independent operators, $K_j^\dag K_{j^\prime}=\sum_{(i,i^\prime)\in\SC}T^{ii^\prime}_{jj^\prime}K_i^\dag K_{i^\prime}$, for $(j,j^\prime)\in\SC^\prime$, where $T$ is a full-rank matrix whose columns are indexed by $(j,j^\prime)$ and rows by $(i,i^\prime)$. This changes ${\bf Q}_{_A}\rightarrow{\bf Q}^\prime_{_A}={\bf Q}_{_A}T$, so while the nullspace of ${\bf Q}_{_A}^\prime$ differs from that of ${\bf Q}_{_A}$, the dimensions of the two nullspaces are identical. Since our results depend only on this dimension, see Theorem~\ref{thm1}, we see that our claim is justified.} In similar fashion, one can use this approach to obtain all later measurements, thus designing a full LOCC protocol for $\EC$ whenever possible. This design approach is described in detail in \cite{myLOCCbyFirstMeas} for the case of implementing a quantum measurement rather than a channel. For measurements, the set of Kraus operators is fixed and unitary $V$ in Eq.~\eqref{eqn22} is equal to the identity; then the free parameters are the $\hat N$ coefficients $c_j$, rather than the $|\SC|$ coefficients $\tilde\CC_{ii^\prime}$. Thus, we see that the approach described in \cite{myLOCCbyFirstMeas} can be directly taken over to the case of channels simply by increasing the dimension of the domain of matrix ${\bf Q}_{_A}$ from ${\hat N}$ \cite{myLOCCbyFirstMeas} to $|\SC|\le{\hat N}^2$, where ${\hat N}$ is the number of Kraus operators defining the measurement (in the first case) or representing the channel (in the second case). 

By Eq.~\eqref{eqn20}, there is always at least one solution of Eq.~\eqref{eqn28}, which we will denote as $\vec c_I$ corresponding to $\AC=I_A$ in Eq.~\eqref{eqn29}. If this is the only solution, then Alice cannot measure first. If this is true for every party, then no LOCC protocol exists for the given channel.

Let us augment ${\bf Q}_{_A}$ by adding the single row $\vec c_I^{\,\dag}$ (and, hoping not to cause confusion, we continue to call this ${\bf Q}_{_A}$). Then $\vec c_I$ is no longer in the nullspace of ${\bf Q}_{_A}$, and we have the following theorem.
\begin{thm1}\label{thm1}
For a given multipartite quantum channel $\EC$, if for every party $\alpha$, the nullspace of ${\bf Q}_\alpha$ is empty, then $\EC$ cannot be implemented by LOCC, even when using an infinite number of rounds.
\end{thm1}
\noindent Note that this theorem provides a way to easily prove certain channels are not LOCC, requiring only the solution of a set of linear equations. Below is a list of examples for which this method provides such a proof. For each example, we give the ratio of the smallest eigenvalue of ${\bf Q}_\alpha^\dag{\bf Q}_\alpha\ge0$ to the largest, minimized over parties $\alpha$, denoting this ratio as $\hat\lambda$.\footnote{The computational complexity of finding all eigenvalues of an $n\times n$ matrix in the large-$n$ limit is ${\cal O}({n^3})$ by Householder reduction to tridiagonal form followed by either QR factorization or direct solution of the resulting characteristic equation \cite{NumRec1}. We believe this calculation is the limiting step in our procedure of checking LOCC-impossibility. I have written Matlab code implementing this procedure, using Matlab's `eig' function to find the eigenvalues, it being plenty fast enough for the problems we have addressed here.} When $\hat\lambda\ne0$, the nullspace of ${\bf Q}_\alpha$ is empty for all parties $\alpha$, and then Theorem~\ref{thm1} shows that the given channel cannot be implemented by LOCC. If $\hat\lambda$ is small, then ${\bf Q}_\alpha$ is close to singular, implying there is a first measurement that introduces only a small error into the protocol. For example, by a small change in the set of Kraus operators one could alter ${\bf Q}_\alpha$ so that it becomes singular. However, it should be remembered that this is just the first measurement of a protocol, so that small $\hat\lambda$ does not by itself indicate there is a protocol that closely approximates the desired channel. In addition, even when $\hat\lambda$ is large, it may still be that the channel lies in the set closure, $\overline{\textrm{LOCC}}$ (at least, we as yet have no argument to the contrary).

Let us now look at the examples.
\begin{enumerate}
  \item\label{itm1} The bipartite channel determined by Kraus operators equal to projectors onto the four maximally entangled Bell states \cite{BellStates}. We find that $\hat\lambda=1$, implying that neither party can make a first measurement and showing that this channel is not LOCC.
  \item\label{itm2} The channel defined by Kraus operators equal to the nine projectors onto the (product) domino states of \cite{Bennett9}, which first demonstrated nonlocality without entanglement. We find that $\hat\lambda=1/6$ for this case, implying that neither party can make a first measurement and showing (the well-known result) that this channel is not LOCC.
  \item \label{itm3}The channel defined by the rotated domino states \cite{Bennett9,ChildsLeung}. We find that for these channels, neither party can measure first, other than in the exceptional cases where at least one pair of the dominos is not rotated at all, in which case there exists a straightforward LOCC protocol for implementing these channels. As seen in Fig.~\ref{fig1} (see Appendix~\ref{app1} for further explanation), $\hat\lambda$ is nonzero, but smoothly approaches zero as the degree of rotation (measured here by $\theta_{\min}$) vanishes. Thus, we see a smooth approach to the existence of a first local measurement, and indeed to a fully LOCC channel that can be implemented using three rounds of communication in this limit, as described in detail in Appendix~\ref{app1}.
\begin{figure}[h]
\includegraphics[scale=0.8]{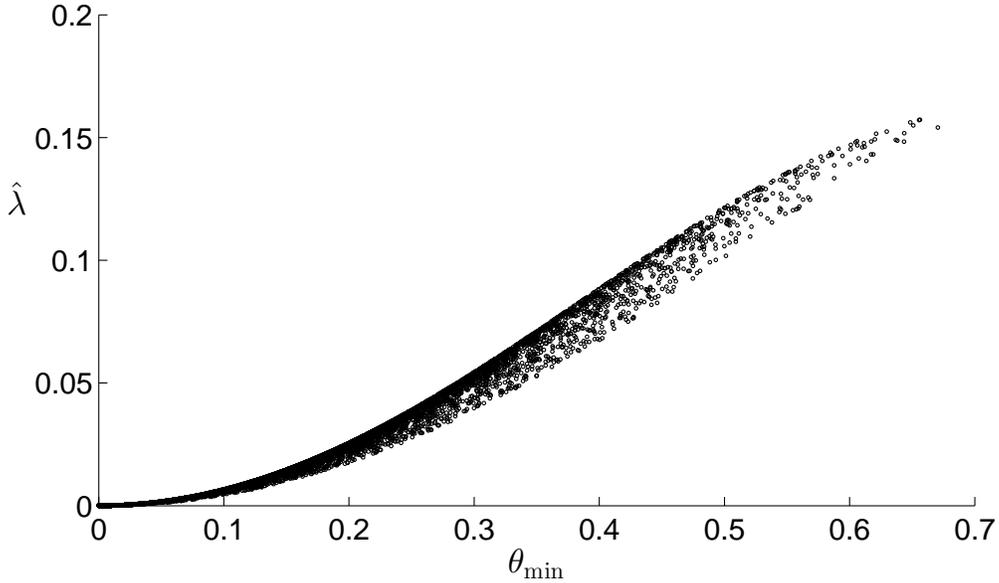}
\caption{\label{fig1}Impossibility of a first measurement for rotated domino states.}
\end{figure}
  \item\label{itm4} Random unitary channels, where the Kraus operators are proportional to (randomly chosen) unitaries. The results in this case depend strongly on the number of random unitaries, $N_u$. For each number of parties, each set of local dimensions, each $N_u$, and equal \emph{a priori} probabilities of the random unitaries, we have numerically checked $100$ different sets of randomly chosen unitaries for each of the following cases: two qubits, three qubits, two qutrits, and a bipartite qubit-qutrit system. In all cases, we have found that no party can ever measure first when $N_u<D$, where $D$ is the dimension of the global Hilbert space $\HC$, while a first measurement is not excluded for any party when $N_u>D$. In these cases, $\hat\lambda$ decreases from no smaller than about $0.01$ when $N_u=2$ to no smaller than $3\times10^{-4}$ when $N_u=D-1$. When $N_u>D$, $\hat\lambda$ is always less than about $10^{-16}$. Thus we see a sharp transition around $N_u=D$. When $N_u=D$,  the situation is more diverse. For three qubits none of the parties can measure first when $N_u=D=8$, $\hat\lambda$ never being less than $10^{-4}$. The case $N_u=D$ yields $\hat\lambda$ ranging between about $10^{-5}$ to $10^{-9}$ for two qutrits and between about $10^{-3}$ to $10^{-8}$ for two qubits, showing that neither party can measure first for these channels, but it is sometimes possible they are close to channels where one or the other party can. For the qubit-qutrit case and $N_u=D=6$, the party with the qubit can never measure first ($\hat\lambda$ is never less than $10^{-5}$), but it may be possible that the party holding the qutrit can ($\hat\lambda$ is always less than about $10^{-16}$). This provides evidence that `almost' all random unitary channels having small $N_u$ are not LOCC.
  \item\label{itm5} The two-qubit channel defined by the five Kraus operators $\ket{n}\bra{\pi_n}$, where $\ket{\pi_1}\propto\ket{0}\left(\alpha_1\ket{0}+\beta_1\ket{1}\right)$, $\ket{\pi_2}\propto\ket{0}\left(\alpha_1\ket{0}-\beta_1\ket{1}\right)$, $\ket{\pi_3}\propto\left(\alpha_3\ket{0}+\beta_3\ket{1}\right)\ket{0}$, $\ket{\pi_4}=\ket{1}\ket{1}$, and $\ket{\pi_5}=\left(\mu\ket{0}+\nu\ket{1}\right)\ket{0}$. These operators provide the optimal (global and separable) measurement for unambiguous state discrimination of the states given in Eq.~($12$) of \cite{myExtViol2}. Coefficients $\alpha_j,\beta_j,\mu,\nu$ are constrained as described in \cite{myExtViol2} and summarized here in Appendix~\ref{app2}, but these constraints leave a continuous class of example channels. It was shown in that paper that this optimal measurement cannot be achieved by finite-round LOCC, but it was not previously known whether or not it could be achieved by LOCC using an infinite number of rounds. We have generated several thousand examples with the coefficients chosen at random, finding that no party can measure first except in the limiting cases of $|\alpha_1|\to0,~|\alpha_1/\beta_1|\to1$, and $\alpha_3\to0$, where LOCC protocols, requiring only a single round of communication, do exist. In all other cases, $\hat\lambda$ is nonzero, but approaches zero as any one of these limiting cases is approached (see Figs.~$2$ and $3$ in Appendix~\ref{app2}). This provides evidence that almost all channels (also measurements) in this class are not LOCC, even with an infinite number of rounds, and the possibility remains that none of them are.
Thus, we have a sequence of channels (also measurements) where no member of that sequence is LOCC, even with an infinite number of rounds, but the limiting channel (measurement) is one-round LOCC. This provides numerical evidence that the set of channels (measurements) that are not LOCC is not closed and that these one-round LOCC channels are on the boundary of LOCC. 
\end{enumerate}

In conclusion, we have adapted the method of \cite{myLOCCbyFirstMeas} for designing LOCC protocols implementing quantum measurements so that it can be used for implementation of quantum channels. We have shown that one can often prove LOCC-impossibility of a quantum channel by solving a set of linear equations and have provided several examples. These examples include a new result suggesting that the class of optimal measurements for unambiguous state discrimination given in \cite{myExtViol2} are all impossible by LOCC, including with an infinite number of rounds, and also a result suggesting that almost all random unitary channels with few unitaries are not LOCC. Additionally, we have obtained numerical evidence that the set of quantum channels that are not LOCC is not closed, and that there exist one-round and three-round LOCC channels that are on the boundary of the set of all LOCC channels. While our results do apply to infinite-round LOCC protocols, they do not apply to the closure of LOCC, and we are presently at work studying ways to include the entire boundary of LOCC in the analysis.

Finally, we would like to point out that our approach to LOCC-impossibility of quantum channels can be recast in the form of semidefinite programs \cite{SDP1}, see Appendix~\ref{app3}. This and other extensions of the results presented here will be discussed elsewhere.

\noindent\textit{Acknowledgments} --- We wish to thank Vlad Gheorghiu for helpful comments on this paper. This work has been supported in part by the National Science Foundation through Grant No. 1205931.

\appendix
\section{Rotated domino states}\label{app1}
The quantum channel of example $3$ of the main text can be represented by a set of Kraus operators that are projectors onto the nine rotated domino states, which are,
\begin{align}\label{eqn520}
\ket{\Psi_1} & = \ket{1}\otimes\ket{1},\notag\\
\ket{\Psi_2} & = \ket{0}\otimes\left(\cos\theta_1\ket{0}+\sin\theta_1\ket{1}\right),\notag\\
\ket{\Psi_3} & = \ket{0}\otimes\left(\sin\theta_1\ket{0}-\cos\theta_1\ket{1}\right),\notag\\
\ket{\Psi_4} & = \ket{2}\otimes\left(\cos\theta_2\ket{1}+\sin\theta_2\ket{2}\right),\notag\\
\ket{\Psi_5} & = \ket{2}\otimes\left(\sin\theta_2\ket{1}-\cos\theta_2\ket{2}\right),\\
\ket{\Psi_6} & = \left(\cos\theta_3\ket{1}+\sin\theta_3\ket{2}\right)\otimes\ket{0},\notag\\
\ket{\Psi_7} & = \left(\sin\theta_3\ket{1}-\cos\theta_3\ket{2}\right)\otimes\ket{0},\notag\\
\ket{\Psi_8} & = \left(\cos\theta_4\ket{0}+\sin\theta_4\ket{1}\right)\otimes\ket{2},\notag\\
\ket{\Psi_9} & = \left(\sin\theta_4\ket{0}-\cos\theta_4\ket{1}\right)\otimes\ket{2},\notag
\end{align}
with $0<\theta_n\le\pi/4$. The `domino states' of \cite{Bennett9} are obtained by setting $\theta_n=\pi/4$ for each $n=1,2,3,4$. It is not difficult to see that if any one (or more) of the $\theta_n=0$, then an LOCC protocol exists for this channel; let us look at one example. First define $\ket{x,y,n,+}=\cos\theta_n\ket{x}+\sin\theta_n\ket{y}$ and $\ket{x,y,n,-}=\sin\theta_n\ket{x}-\cos\theta_n\ket{y}$ and suppose $\theta_1=0$ so that $\ket{\Psi_2}=\ket{0}$ and $\ket{\Psi_3}=\ket{1}$. Then this set of states can be perfectly distinguished (and the channel implemented) by the following LOCC protocol. Bob (the second party) begins by measuring with Kraus operators $[0]$ and $[1]+[2]$, where $[\phi]=\ket{\phi}\bra{\phi}$. If he obtains $[0]$, Alice measures with outcomes $[0]$ (identifying $\ket{\Psi_2}$), $[1,2,3,+]$ ($\ket{\Psi_6}$), and $[1,2,3,-]$ ($\ket{\Psi_7}$). Otherwise, Alice measures with outcomes $[0]+[1]$ and $[2]$. If she obtains $[2]$, Bob measures with outcomes $[1,2,2,+]$ ($\ket{\Psi_4}$) and $[1,2,2,-]$ ($\ket{\Psi_5}$), and any other outcomes have zero probability of occurring. Otherwise, he measures with outcomes $[1]$ and $[2]$ ($[0]$ now has zero probability), followed by Alice measuring $[0]$ ($\ket{\Psi_3}$) and $[1]$ ($\ket{\Psi_1}$) for his first outcome and $[0,1,4,+]$ ($\ket{\Psi_8}$) and $[0,1,4,-]$ ($\ket{\Psi_9}$) for his second. This completes the protocol, which involves three rounds of communication.

When $\theta_n\ne0$ for all $n$, the method in the main text of finding the nullspace of $\bf{Q}_\alpha$ for parties $\alpha$ shows that these channels cannot be implemented by LOCC. In Fig.~$1$ of the main text, we have plotted $\hat\lambda$ vs. $\theta_{\min}=\min{\left(\theta_1,\theta_2,\theta_3,\theta_4\right)}$ for each of $10,000$ randomly chosen sets of the $\theta_n$, illustrating the smooth approach to LOCC where $\hat\lambda\rightarrow0$ as any one of the $\theta_n\rightarrow0$.

\section{Class of unambiguous state discrimination examples from \cite{myExtViol2}}\label{app2}
Example $5$ of the main text concerns the unambiguous discrimination of the following set of four linearly independent states on two qubits, $\SC=\{\eta_j,\ket{\Phi_j}\}$, each given with \emph{a priori} probability $\eta_j$,
\begin{align}\label{eqn34}
|\Phi_1\rangle & = \frac{1}{\sqrt{\left|\beta_3\right|^2+\left|\alpha_3\beta_1\right|^2}}\left(\beta_3^\ast\beta_1^\ast\ket{00}+\beta_3^\ast\alpha_1^\ast\ket{01}-\alpha_3^\ast\beta_1^\ast\ket{10}\right),\notag\\
|\Phi_2\rangle & = \frac{1}{\sqrt{\left|\beta_3\right|^2+\left|\alpha_3\beta_1\right|^2}}\left(\beta_3^\ast\beta_1^\ast\ket{00}-\beta_3^\ast\alpha_1^\ast\ket{01}-\alpha_3^\ast\beta_1^\ast\ket{10}\right),\notag\\
|\Phi_3\rangle & = \ket{10},\notag\\
|\Phi_4\rangle & = \ket{11},
\end{align}
where for $j=1,3$ we require that $\alpha_j\ne0$, $\beta_j\ne0$, $\left|\alpha_j\right|^2+\left|\beta_j\right|^2=1$, $|\alpha_1|<|\beta_1|$, and $\eta_1=\eta_2$ (the penultimate of these constraints was recorded incorrectly as $|\alpha_1|\le|\beta_1|$ in \cite{myExtViol2}). Finally, we apply one additional restriction to these coefficients, which ensures there is a unique measurement that is optimal for unambiguously discriminating these states. This final restriction is,
\begin{align}\label{eqn49}
\left(1-\left|\frac{\alpha_1\beta_3}{\beta_1}\right|^2\right)^2\ge\frac{\eta_3}{4\eta_1}\left|\frac{\alpha_3}{\beta_1}\right|^2\left(|\beta_3|^2+|\alpha_3\beta_1|^2\right).
\end{align}
These restrictions leave a wide range of allowed values for the coefficients.

The quantum channel for this example may be described by the set of five Kraus operators,
\begin{align}\label{eqn35}
K_n=\sqrt{p_n}\ket{n}\bra{\Psi_n},
\end{align}
where
\begin{align}\label{eqn36}
|\Psi_1\rangle & =  q\ket{0}\left(\alpha_1\ket{0}+\beta_1\ket{1}\right),\notag\\
|\Psi_2\rangle & = q\ket{0}\left(\alpha_1\ket{0}-\beta_1\ket{1}\right),\notag\\
|\Psi_3\rangle & = \left(\frac{\alpha_3}{\beta_3}\ket{0}+\ket{1}\right)\ket{0},\notag\\
|\Psi_4\rangle & = \ket{11},
\end{align}
with $p_4=1$ and
\begin{align}\label{eqn41}
p_1&=p_2=\frac{1}{2\left|q\beta_1\right|^2},\notag\\
p_3&=\frac{|\beta_3|^2\left(1-\left|\alpha_1/\beta_1\right|^2\right)}{\left(1-\left|\alpha_1\beta_3/\beta_1\right|^2\right)},\notag\\
q&=\frac{\sqrt{\left|\beta_3\right|^2+\left|\alpha_3\beta_1\right|^2}}{2\alpha_1^\ast\beta_1^\ast\beta_3^\ast}.
\end{align}
Finally, $K_5=\ket{5}\bra{\Psi_5}$ with $\ket{\Psi_5}=\ket{\pi_5}\otimes\ket{0}$, and
\begin{align}\label{eqn42}
\ket{\pi_5}=\mu\ket{0}+\nu\ket{1},
\end{align}
where
\begin{align}\label{eqn43}
\mu&=\sqrt{1-\left|\alpha_1/\beta_1\right|^2-\left|\alpha_3/\beta_3\right|^2 p_3},\notag\\
\nu&=-e^{i\phi}\sqrt{1- p_3},
\end{align}
and $-\phi$ must be the same as the phase of $\alpha_3/\beta_3$. The foregoing choices of the $p_j$ yield the unique, optimal measurement for unambiguous discrimination of the states in \eqref{eqn34}, which is a manifestly separable measurement, as all POVM elements are product operators. Hence, separable measurements are as good as global ones for this task.

Note, for example, that when $\alpha_3=0$, this channel can be implemented by Alice measuring in the $\ket{0},\ket{1}$ basis, after which Bob can measure as follows: if Alice obtained $\ket{1}$, Bob measures with outcomes $\ket{3}\bra{0}$ and $\ket{4}\bra{1}$, while if Alice obtained $\ket{0}$, he measures with outcomes proportional to $\ket{1}\left(\alpha_1^\ast\bra{0}+\beta_1^\ast\bra{1}\right)$, $\ket{2}\left(\alpha_1^\ast\bra{0}-\beta_1^\ast\bra{1}\right)$, and $\ket{5}\bra{0}$. Thus, these special cases are one-way LOCC. The cases $\alpha_1=0$ and $\alpha_1=\beta_1$ are also LOCC.

\begin{figure}[h]
\includegraphics[scale=0.8]{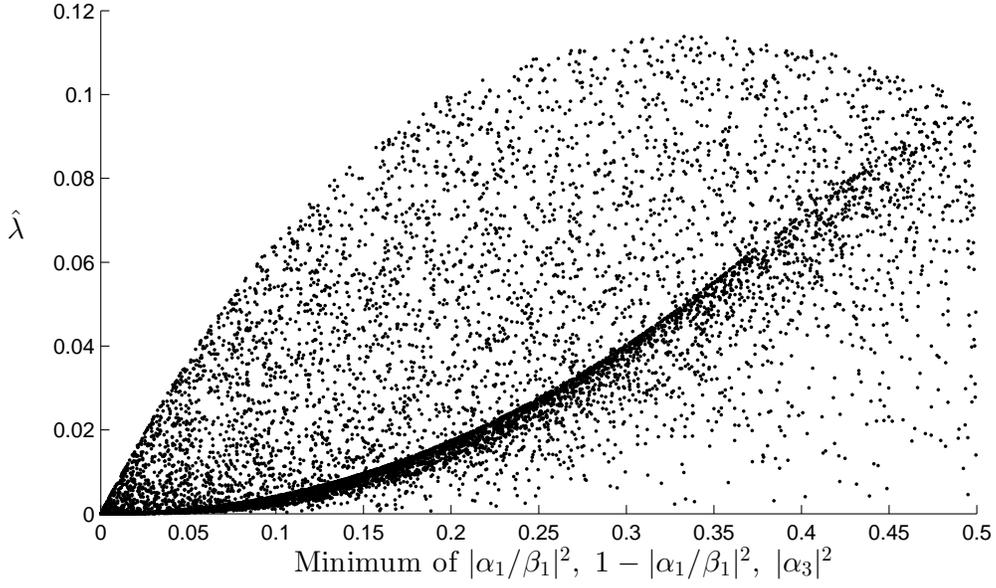}
\caption{\label{fig2}Impossibility of a first measurement for the unambiguous state discrimination examples from \cite{myExtViol2}. }
\end{figure}

\begin{figure}[h]
\includegraphics[scale=0.8]{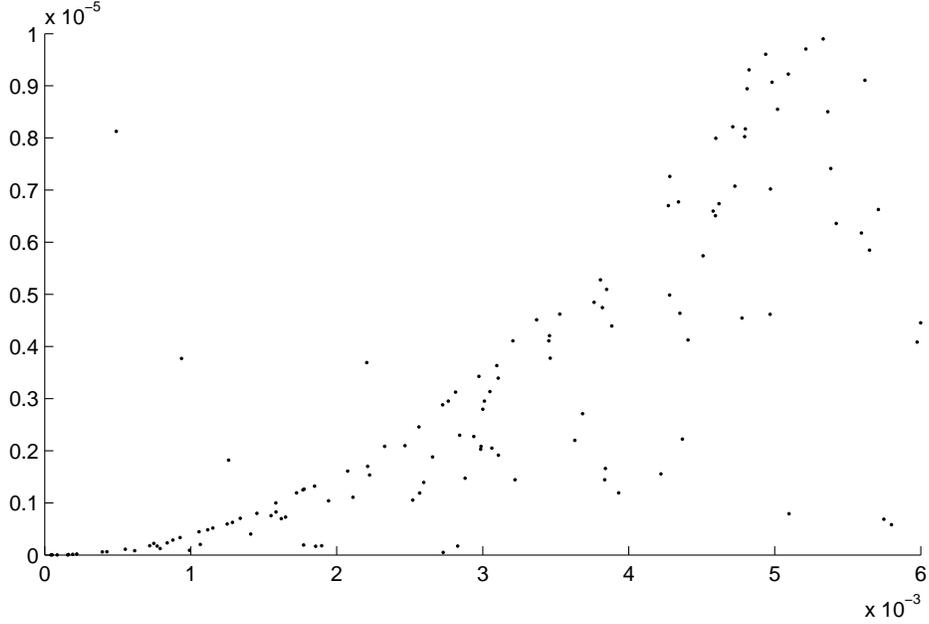}
\caption{\label{fig3}Impossibility of a first measurement for the unambiguous state discrimination examples from \cite{myExtViol2}. A closer look near the origin of Fig.~\ref{fig2}.}
\end{figure}

We next ask whether the channel described by Kraus operators $K_n$ can be achieved by LOCC. We have randomly generated $10,000$ instances of these channels and checked the nullspace of matrices ${\bf Q}_\alpha$ for each instance. We find that for each such instance, no party can measure first in an LOCC protocol implementing the given channel. We also observe a smooth approach to LOCC channels at $\alpha_1=0,~\alpha_1=\beta_1$ or $\alpha_3=0$. Figs.~\ref{fig2} and \ref{fig3} illustrate these results.

\section{Semidefinite Programming}\label{app3}
Let us now see how the approach of the main text to prove LOCC-impossibility of a quantum channel can be reformulated as a set of semidefinite programs \cite{SDP1}. Instead of reshaping the mathematical objects in Eq.~($6$) of the main text, we may follow a somewhat different approach. If Alice can measure first then $\AC\not\propto I_A$ and with
\begin{align}\label{eqn26}
\RC_\omega:=d_{\bar A}\pTr{A}{\Lambda_\omega^\dag\AC}=\sum_{i,i^\prime=1}^N\CC_{ii^\prime}\Tr{\left[\Lambda_\omega^\dag\otimes I_{\bar A}\right]K_i^\dag K_{i^\prime}}=\sum_{i,i^\prime=1}^N\CC_{ii^\prime}\QC_{i^\prime i}^{(\omega0)},~\omega=1,\cdots,d_A^2-1,
\end{align}
$\RC_\omega\ne0$ in Eq.~\eqref{eqn26} for at least one value of $\omega\ne0$. In addition, for each $\AC\not\propto I_A$ satisfying this condition, there is also $\AC^\prime=I_A-\AC\not\propto I_A$ which satisfies this same condition and with the same nonzero value of $\RC_\omega$, apart from an extra minus sign. The reason this is true is that if $\AC\otimes I_{\bar A}=\sum_{j=1}^N c_j K_j^{\prime\dag} K_j^\prime$, then $\left(I_A-\AC\right)\otimes I_{\bar A}=\sum_{j=1}^N\left(1-c_j\right)K_j^{\prime\dag} K_j^\prime$ with $1-c_j\ge0$ (which one can show must also hold), and using $\Tr{\Lambda_\omega^\dag}=0$ when replacing $\AC$ by $I_A-\AC$ in Eq.~\eqref{eqn26}, we see that everything follows for $\left(I_A-\AC\right)$ just as it did for $\AC$, apart from the replacement of $\RC_\omega\rightarrow-\RC_\omega$ for each $\omega$. This means that the minimum of $\RC_\omega$ is strictly less than zero for at least one value of $\omega\ne0$ unless Alice \emph{cannot} measure first. In the latter case, the only allowed operators $\AC$ in Eq.~($5$) of the main text are proportional to $I_A$, and then $\RC_\omega=0$ for all $\omega\ne0$.

Note that defining matrix $\mathbf{C}=\textrm{diag}(\{c_j\})$, we have from Eq.~($4$) in the main text that $\CC = V^\dag\mathbf{C}V\ge0$. That is, $\CC$ is a positive semidefinite matrix. Similarly, $I_N-\CC\ge0$, with $I_N$ the $N\times N$ identity matrix. Let us define
\begin{align}\label{eqn27}
X=\left(\begin{array}{c c}
  \CC & 0\\
 0& I_N-\CC\\
\end{array}\right)\notag\\
M_\omega=\left(\begin{array}{c c}
  \QC^{(\omega 0)} & 0\\
 0& 0\\
\end{array}\right)\\
A_{\mu\nu}=\left(\begin{array}{c c}
  \QC^{(\mu\nu)} & 0\\
 0& 0\\
\end{array}\right)\notag
\end{align}
and
\begin{align}\label{eqn728}
B_{ij}=
\begin{dcases}
    \ket{j}\bra{i}+\ket{j+N}\bra{i+N}, & i,j=1,\cdots,N\\
    \ket{j}\bra{i},&i=1,\cdots,N;~j=N+1,\cdots,2N\text{ or }j=1,\cdots,N;~i=N+1,\cdots,2N\\
	0,&\text{ otherwise}
\end{dcases}
\end{align}
Then, we have the following theorem involving a set of semidefinite programs \cite{SDP1}, which supplies another means of proving that a given quantum channel \emph{cannot} be implemented by LOCC.
\begin{customthm}{S1}\label{eight}
For any quantum channel $\EC$, consider the following set of semidefinite programs, one for each $\omega=1,\cdots,d_A^2-1$.
\begin{align}\label{eqn729}
&\min_X\Tr{XM_\omega}\notag\\
\textrm{subject to }~&\Tr{XA_{\mu\nu}}=0,~\mu=0,\cdots,d_A^2-1,~\nu=1,\cdots,d_{\bar A}^2-1,\notag\\
					&\Tr{XB_{ij}}=
					\begin{dcases}
							\delta_{ij},&i,j=1,\cdots,N,\\
							0,&\text{ otherwise}
					\end{dcases}\\
				&X\ge0\notag
\end{align}
where $X,M_\omega,A_{\mu\nu}$, and $B_{ij}$ are defined above and $\QC^{(\mu\nu)}$ are defined in terms of Kraus representation $\{K_i\}$ of $\EC$ by Eq.~($7$) of the main text. If the solution to each of these semidefinite programs is equal to zero, then Alice cannot measure first to initiate an LOCC protocol implementing $\EC$. Considering an analogous set of semidefinite programs for each of the parties, then if no party can measure first, channel $\EC$ cannot be implemented by LOCC, even with an infinite number of rounds of communication.
\end{customthm}

It has been pointed out to us that since it is necessary that the solution to these (primal) SDP's be exactly zero, this is a computationally untenable approach to the problem. However, it may be that their dual programs can be solved analytically in some cases, and we are presently studying this possibility.

\end{document}